\documentclass{PHYEAUTH}
\usepackage{graphicx}
\usepackage{amsmath}
\usepackage{amssymb}

\begin{document}

\begin{frontmatter}

\title{
Dynamic nuclear polarization and Knight shift measurements \\
in a breakdown regime of integer quantum Hall effect
}

\author[address1]{Minoru~Kawamura\thanksref{thank1}},
\author[address1]{Hiroyuki~Takahashi},
\author[address1]{Satoru~Masubuchi},
\author[address2]{Yoshiaki~Hashimoto},
\author[address2,address3]{Shingo~Katsumoto},
\author[address1]{Kohei~Hamaya},
\author[address1,address3]{Tomoki~Machida}

\address[address1]{Institute of Industrial Science, 
	University of Tokyo, Tokyo 153-8505, Japan}

\address[address2]{Institute for Solid State Physics, 
	University of Tokyo, Chiba 277-8581, Japan}

\address[address3]{Institute for Nano Quantum Information Electronics, 
	University of Tokyo, Tokyo 153-8505, Japan}

\thanks[thank1]{
Corresponding author.
E-mail: minoru@iis.u-tokyo.ac.jp}

\begin{abstract}
Nuclear spins are polarized electrically
in a breakdown regime of an odd-integer quantum Hall effect (QHE).
Electron excitation
to the upper Landau subband with the opposite spin polarity
flips  nuclear spins through the hyperfine interaction.
The polarized nuclear spins reduce 
the spin-splitting energy and  accelerate the QHE breakdown.
The Knight shift of the nuclear spins is also  measured
by tuning electron density during the irradiation of 
radio-frequency magnetic fields.
\end{abstract}

\end{frontmatter}

Interplay between nuclear spins and electron spins has
recently attracted great interests in the fundamental research of 
electron transport in low-dimensional
systems \cite{Kronmuller1999,Machida2002,Hashimoto2002,Ono2004}.
In integer quantum Hall (QH) systems,
inter-edge-channel scatterings of electrons accompanied by 
spin flips cause nuclear spins to polarize
via hyperfine interaction\cite{Machida2002}.
In fractional QH systems near spin-polarized/unpolarized
transition, the hyperfine interaction also takes part 
in the dynamic nuclear polarization (DNP)\cite{Kronmuller1999,Hashimoto2002}.

On the other hand, when electron spins are polarized, 
the hyperfine interaction generate an effective magnetic field 
for nuclear spins.
As a result, nuclear magnetic resonance (NMR) frequency 
is shifted depending on the electron spin polarization.
This frequency shift is referred to as the Knight shift.
Since the Knight shift is proportional to the electron spin polarization,
studies on the Knight shift revealed
electron spin properties in two-dimensional electron gases (2DEGs)
that had not been accessed by conventional transport 
measurements;
for example, finite-size skyrmions\cite{Barrett1995}
and domain structures with different spin configuration
in fractional QH systems\cite{Stern2004}.
Thus, a new method for electrical polarization of nuclear spins 
and for detection of the Knight shift
will have a potential to uncover
novel spin-dependent  phenomena in QH systems.


In this paper, 
we report 
electrical polarization of nuclear spins
and all-electric measurements of the Knight shift
in a breakdown regime of odd-integer quantum Hall effect.
When an applied electric current exceeds a critical current $I_{\rm c}$
of QHE breakdown,
electrons are excited to the upper Landau subband
with the opposite spin polarity.
The flip of electron spin $\boldmath{S}$ flops  nuclear spin $\boldmath{I}$
through the hyperfine interaction,
$
\boldmath{H}_{\rm hyp} = A\boldmath{I}\cdot\boldmath{S} = 
A(I^{+}S^{-} + I^{-}S^{+})/2 + AI_{z}S_{z},
$
where $A$ is the hyperfine constant.
As a result, nuclear spins are polarized
along  the external magnetic field $B_{\rm ext}$ ($\langle I_z \rangle > 0$).
The polarized nuclear spins generate an effective magnetic field  $B_{\rm NS}$
for electron spins.
The $B_{\rm NS}$ reduces the spin-splitting energy
$E_{\rm s}$ = $|g|\mu_{\rm B}(B_{\rm ext} - B_{\rm NS})$,
where $g$ is the effective $g$-factor and $\mu_{\rm B}$ is the Bohr magneton.
Thus, the DNP accelerates the QHE breakdown.
Since  the DNP is induced in the bulk part of the 2DEG in this method,
the DNP can be utilized as a probe for Knight shift measurements
in the bulk part of QH conductor.


Experiments were performed using
a 20-$\mu$m-wide Hall-bar device
fabricated from a wafer of GaAs/Al$_{0.3}$Ga$_{0.7}$As single heterostructure
with 2DEG. 
The mobility and sheet carrier density of the 2DEG
are 220 m$^{2}$/Vs and 1.53 $\times$ 10$^{15}$ m$^{-2}$ at 4.2 K,
respectively.
Gate voltage $V_{\rm G}$ is applied to the front gate electrode
to tune the carrier density of the 2DEG.
All the measurements were performed in a dilution 
refrigerator with a base temperature of 30 mK.
A single-turn coil around the device was used to irradiate
radio-frequency (rf) magnetic fields.

\begin{figure}[t]
\begin{center}\leavevmode
\includegraphics[width=\linewidth]{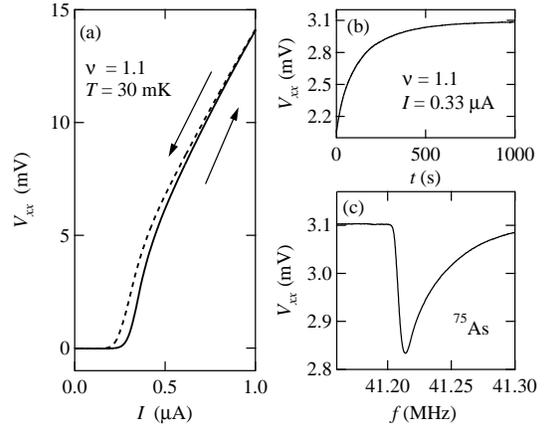}
\caption{
	(a) $V_{xx}$-$I$ characteristics at $\nu$ = 1.1
	taken by sweeping current in positive (solid) 
	and negative (dashed) directions.
	(b) Time evolution of $V_{xx}$ after switching current from $I$ = 0
	to 0.33 $\mu$A at $t$ = 0.
	(c) NMR spectrum for $^{75}$As detected by measuring $V_{xx}$.
}
\label{Hallbar}
\end{center}
\end{figure}

Figure~\ref{Hallbar}(a) shows the voltage-current ($V_{xx}$-$I$)
characteristics at a Landau level filling factor
under the gate $\nu_{\rm G}$ = 1.1  
($B_{\rm ext}$ = 3.21 T and $V_{\rm G}$ = $-$0.33 V).
The solid and dashed curves are respectively
obtained by sweeping the current in positive and negative directions.
The critical current $I_{\rm c}$ = 0.24 $\mu$A in the down-sweep curve 
is smaller than $I_{\rm c}$ = 0.29 $\mu$A in the up-sweep curve.
Figure~\ref{Hallbar}(b) shows the time evolution of $V_{xx}$
at $\nu_{\rm G}$ = 1.1 after switching bias current 
from $I$ = 0 $\mu$A to 0.33 $\mu$A.
The value of $V_{xx}$ increases slowly with a long relaxation time over 500 s.
After complete saturation of $V_{xx}$ at $I$ = 0.33 $\mu$A,
rf-magnetic field parallel to the 2DEG is applied.
The value of $V_{xx}$ decreases at the NMR frequency
of $^{75}$As as shown in Fig.~\ref{Hallbar}(c).

The slow time evolution in $V_{xx}$ and
the detection of the NMR signal definitely show that the nuclear 
spins are dynamically polarized in the QHE breakdown regime,
similar to our earlier work using different devices 
without gate electrode\cite{Kawamura2007}.
The shift of $I_{\rm c}$ toward the smaller-current side
in Fig.~\ref{Hallbar}(a) and  the increase in $V_{xx}$ in Fig.~\ref{Hallbar}(b)
indicate the acceleration of the QHE breakdown by the DNP,
which can be understood by 
the reduction of the spin-splitting energy $E_{\rm s}$.
These results show  $\langle I_{z} \rangle > 0$.

Let us turn our attentions to the Knight shift measurement.
Recently all-electric detection of the Knight shift was demonstrated
using QH edge channels\cite{Masubuchi2006}.
We employed a similar technique to measure the Knight shift
in the bulk part of the QH system.
First, nuclear spins are polarized 
at $\nu_{\rm DNP} = 1.1$ ($B_{\rm ext}$ = 3.21 T)
by applying pumping current $I_{\rm pump}$ = 0.33 $\mu$A.
Next, the current is turned off and the fillng factor
is set to $\nu_{\rm rf}$ by changing $V_{\rm G}$.
Then, the rf-magnetic field with a frequency $f$ is applied for 5 sec.
Finally, the filling factor is set back to $\nu_{\rm DNP}$ = 1.1 and 
the time evolution of $V_{xx}$ is recorded.
By repeating the above procedure with different $f$,
an NMR spectrum ($V_{xx}$-$f$ curve) is obtained.

\begin{figure}[t]
\begin{center}\leavevmode
\includegraphics[width=\linewidth]{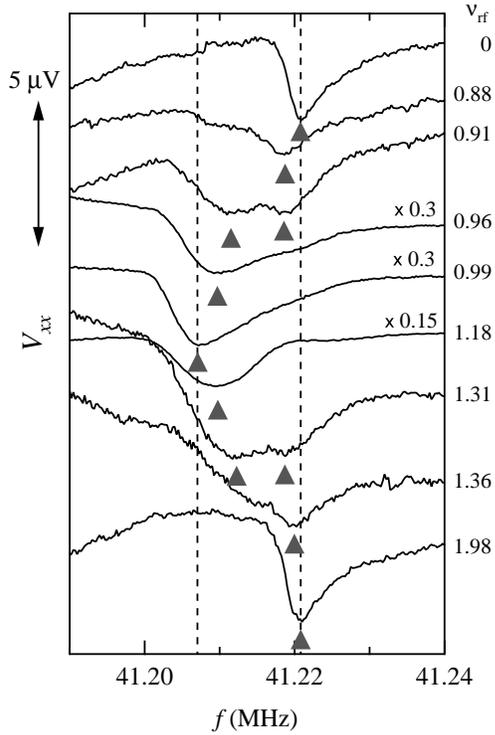}
\caption{
NMR spectra taken by applying rf-magnetic field
at various filling factors between 0 and 2.
The positions of the dips are indicated by triangles.
The dashed lines show  NMR frequencies for
$\nu_{\rm rf}$ = 0 and 0.99.
The traces are shifted vertically for clarity.
} \label{Knightshift}
\end{center}
\end{figure}

We carry out the above procedure at various $\nu_{\rm rf}$
between $\nu_{\rm rf}$ = 0 and 2,
and the NMR spectra of $^{71}$Ga are obtained
as shown in Fig.~\ref{Knightshift}.
The NMR frequencies are different from each other
depending on $\nu_{\rm rf}$.
In the case of $\nu_{\rm rf}$ = 0 (depletion),
the nuclear spins are not affected by electron spins.
Therefore the NMR frequency for $\nu_{\rm rf} = 0$ is equal to 
the intrinsic NMR frequency of $^{71}$Ga without Knight shift.
In the cases of  $\nu_{\rm rf}$ $\neq$ 0, 
the NMR frequencies depend on the electron spin polarization
because the nuclear spins interact with electrons 
in QH states of $\nu_{\rm rf}$ during rf-field application.
In the NMR spectrum for $\nu_{\rm rf}$ = 0.99, 
where the electron spins are fully polaried,
the resonance dip appears at $f$ = 40.2067 MHz.
The frequency gives the largest shift 14.3 kHz 
from the NMR frequency $f$ = 40.2210 MHz for $\nu_{\rm rf}$ = 0.
In the spectrum for $\nu_{\rm rf}$ = 1.98,  where both up-spin and down-spin 
subbands of the lowest Landau level are equally occupied,
the Knight shift of the 
NMR frequency is almost zero.
We infer that the values of the Knight shift are reasonable
as compared with the Knight shift reported 
in  earlier works\cite{Barrett1995,Stern2004,Masubuchi2006}.

At intermediate filling factors $\nu_{\rm rf}$ = 0.91 and 1.31,
the NMR spectra exhibit double-dip structures.
The higher frequency dips appear at frequencies 
close to the NMR frequency for $\nu_{\rm rf}$ = 0.
The NMR spectra with two resonant frequencies recall
the "dispersive-line-shape spectrum" observed in quantum Hall states
near $\nu$ = 0.9 and 1.1\cite{Desrat2002,Tracy2006}.
Although the physical interpretation of the double-dip structure
is not known yet,
it suggest the coexistence of electron systems
with different spin polarization.

\section*{Acknowledgements}
This work is supported by the Grant-in-Aid from MEXT
and  Special Coordination Funds for Promoting Science and Technology.

\end{document}